# Foreign exchange risk premia: from traditional to state-space analyses

Siwat Nakmai*
15 April 2016

**Abstract**

This paper examines foreign exchange risk premia from simple univariate regressions to the state-space method. The adjusted traditional regressions properly figure out the existence and time-evolving property of the risk premia. Successively, the state-space estimations overall are quite rationally competent in examining the essence of time variability of the unobservable risk premia. To be more precise, the coefficients on the lagged estimated time-series are significant and the disturbance combined from the observation and transition equations in the state-space system, rational and premium errors, respectively, is statistically white noise. Such the two residuals are discovered to move oppositely with their covariance approaching zero suggested by the empirics. Besides, foreign exchange risk premia are projected and found significantly stationary at level and relatively volatile throughout time with some clustering. This volatility is however not quite dominant in the deviations of forward prediction errors.

## 1. Introduction

In foreign exchange markets, there exist risk premia once forward rates differ from anticipations with the same horizon. Nevertheless, the risk premia are unobserved due to unobservable expectations. Hypothetically, they are temporally inherent in the linear dynamic system of forward pricing. To build up the system, it can be in a state-space representation, accounting for the unobservability and changeability in time of the variable of interest, which is the unseen risk premia in this paper. To proceed a risk-premium state-space analysis, it may be more convincing to begin at a foundation. Thus, to do so, it is perhaps to understand and model the risk premia steadily from fundamentals, which are traditional-based univariate regressions, progressively to a generalized state-space approach with its Kalman-filter algorithm and log likelihood function. Accordingly, this risk-premium-analytic paper is outlined as background, theoretical framework, data descriptions and statistics, and empirics, respectively.

* Certified FRM, currently PhD economics candidate at Catholic University of Milan, Italy.
E-mail: siwat.nakmai@unicatt.it

# 2. Background

In studying foreign exchange risk premia, Fama (1984) employed univariate regressions to explore the existence of time-variant risk premia. Later, Wolff (1987) identified and measured the risk premia in forward valuation via state-space constructions. He found that the state-space methodology is operative in grabbing the core of the time-series characters of the risk premia, which are also time persistent and prevailing in forward-error fluctuations. As Wolff (1987) simply and theoretically imposed that the covariance of rational and premium errors is zero, Cheung (1993) relaxed this assumption and successfully discovered that such non-zero covariance improved his modeling as obtaining larger maximum log likelihood values. He showed that the risk premia are stationary and highly persistent and the covariations between risk premia and unexpected currency changes are negative. Along these, this paper attempts to go over foreign-exchange-risk-premia analyses stage by stage. In doing so, Fama-related (1984) univariate regressions are done prior to conducting studies similar to Wolff (1987) and Cheung (1993). Pertaining to the sample, it is maybe better to take in hand currencies with their forward dynamics that have been significantly natural as long as possible up to the present time (i.e. almost fully floating and having considerably lengthy data series). With the US dollar (USD) as the domestic currency as usual, the foreign exchanges selected are one that is experienced, which is the Great British pound (GBP), and the other that is promising, which is the Hong Kong dollar (HKD). The former has been very well established in forwards market since January 1979. For the latter, Hong Kong is known as one of the largest financial hubs in Asia while its currency forwards have been more newly recognized since January 1996. Both are also considered ones of the most liquid and floating currencies for decades in financial markets globally.[1]

# 3. Theoretical framework

Prior to analyzing foreign exchange risk premia in a state-space scheme, which is initiated upon their unobservable and time-varying characteristics, the relevant definitions and traditional-based structure are formerly considered in order to realize such unobservability and time variation.

## 3.1 Traditional-based modeling

Consistent with Fama (1984), recall the fundamental assumptions of foreign exchange market efficiency and rationality and the two univariate regressions on forward-spot differentials, $f_t - s_t$. One is of forward forecast errors, $f_t - s_{t+1}$, and the other is of spot changes, $s_{t+1} - s_t$ (all in natural logarithms). They are (1) and (2), respectively.

[1] http://www.bis.org/publ/rpfx13fx.pdf

$$f_t - s_{t+1} = \alpha_1 + \beta_1(f_t - s_t) + \varepsilon_{1,t+1} \quad (1)$$
$$s_{t+1} - s_t = \alpha_2 + \beta_2(f_t - s_t) + \varepsilon_{2,t+1} \quad (2)$$

where $\varepsilon_{1,t+1} \sim \text{i.i.d.}, N(0, \sigma^2_{\varepsilon_1}), \varepsilon_{2,t+1} \sim \text{i.i.d.}, N(0, \sigma^2_{\varepsilon_2})$,
$t \in \{1, \cdots, T\}$, and,
$(f_t - s_{t+1}) + (s_{t+1} - s_t) = f_t - s_t$ by (1)+(2); hence,
$\alpha_1 + \alpha_2 = 0, \beta_1 + \beta_2 = 1, \varepsilon_{1,t+1} + \varepsilon_{2,t+1} = 0$

Let $rp_t$ be foreign exchange risk premia or forward premia, $rp_t \begin{cases} > 0, \text{premia} \\ < 0, \text{discounts} \end{cases}$
$re_{t+1}$ be rational errors or errors from rational expectation of spots,
$fe_t$ be forward errors or errors from forwards predicting spots,

then $rp_t = f_t - s^e_{t+1}, rp_t$ unobserved since $s^e_{t+1} = E_t s_{t+1}$ is unobserved (3)
$re_{t+1} = s^e_{t+1} - s_{t+1}, re_{t+1} \sim$ white noise due to rational expectations (4)
$fe_t = f_t - s_{t+1} = rp_t + re_{t+1}$ (5)

Accordingly,

$$s_{t+1} - s_t = s^e_{t+1} - s_t - re_t = \Delta s^e_{t+1} - re_{t+1} \quad (6)$$
$$f_t - s_t = rp_t + s^e_{t+1} - s_t = rp_t + \Delta s^e_{t+1} \quad (7)$$

Thus, with (5), (6), and (7) and the white-noise $re_{t+1}$ bringing about that $\sigma(re_{t+1}, rp_t + \Delta s^e_{t+1}) = \sigma(-re_{t+1}, rp_t + \Delta s^e_{t+1}) = 0$, the estimated coefficients ($\beta_1$ and $\beta_2$) are as follows.

$$\begin{aligned} \beta_1 &= \frac{\sigma(rp_t + re_{t+1}, rp_t + \Delta s^e_{t+1})}{\sigma^2(rp_t + \Delta s^e_{t+1})} \\ &= \frac{\sigma(rp_t, rp_t + \Delta s^e_{t+1})}{\sigma^2(rp_t + \Delta s^e_{t+1})} + \frac{\sigma(re_{t+1}, rp_t + \Delta s^e_{t+1})}{\sigma^2(rp_t + \Delta s^e_{t+1})} \\ &= \frac{\sigma^2(rp_t) + \sigma(rp_t, \Delta s^e_{t+1})}{\sigma^2(rp_t + \Delta s^e_{t+1})} \end{aligned} \quad (8)^2$$

$$\begin{aligned} \beta_2 &= \frac{\sigma(\Delta s^e_{t+1} - re_{t+1}, rp_t + \Delta s^e_{t+1})}{\sigma^2(rp_t + \Delta s^e_{t+1})} \\ &= \frac{\sigma(\Delta s^e_{t+1}, rp_t + \Delta s^e_{t+1})}{\sigma^2(rp_t + \Delta s^e_{t+1})} + \frac{\sigma(-re_{t+1}, rp_t + \Delta s^e_{t+1})}{\sigma^2(rp_t + \Delta s^e_{t+1})} \\ &= \frac{\sigma^2(\Delta s^e_{t+1}) + \sigma(rp_t, \Delta s^e_{t+1})}{\sigma^2(rp_t + \Delta s^e_{t+1})} \end{aligned} \quad (9)^2$$

[2] $\sigma(a, a+b) = E[(a - E(a))(a + b - E(a+b))] = E(a^2) - (E(a))^2 + E(a+b) - E(a)E(b) = \sigma^2(a) + \sigma(a,b)$

## Time-varying foreign exchange risk premia

After introducing the traditional univariate regressions related to foreign exchange risk premia, now consider the time-varying characterization of foreign exchange risk premia, $rp_t$. If future spot rates deviate from the corresponding forwards, i.e. $s_{t+1} \neq f_t$, or the variability of risk premia outweighs that of expected spot changes, i.e. $\sigma^2(rp_t) > \sigma^2(\Delta s_{t+1}^e)$ where $rp_t$ and $\Delta s_{t+1}^e$ are the components of $f_t - s_t$ (refer to (7)), there could exist the time evolvement of $rp_t$ significantly inherent in $f_t - s_t$ (Fama, 1984).

$$s_{t+1} \neq f_t \quad (10)$$
$$\sigma^2(rp_t) > \sigma^2(\Delta s_{t+1}^e) \quad (11)$$

(10) and (11) are the conditions for $rp_t$ varying through time in $f_t - s_t$ where $s_{t+1} \neq f_t$ and $\sigma^2(rp_t) > \sigma^2(\Delta s_{t+1}^e)$ can be drawn from $\beta_2 \neq 1$ and $\beta_1 > \beta_2$, respectively.

So, after reviewing the theoretical grounds concerning foreign exchange risk premia, $rp_t$, let me extensively present alternatively simple analyses in order to explore $rp_t$, the variable of interest, more directly. First, deal with $\beta_2 \neq 1$ and, second, with $\beta_1 > \beta_2$.

First, $\beta_2 \neq 1$ equals $\beta_2 - 1 \neq 0$ indicating $-\beta_1 \neq 0$ since $\beta_1 + \beta_2 = 1$ (refer to (1) and (2)). The sign of the t-statistic value for $-\beta_1$ is then basically opposite to that for $\beta_1$; but, the two-tail p-values are just equivalent. Alternatively, simply regress $-fe_t = s_{t+1} - f_t$ on $f_t - s_t$ in order to retrieve the estimator equating to $-\beta_1$.

$$-fe_t = s_{t+1} - f_t = \alpha_3 + \beta_3(f_t - s_t) + \varepsilon_{3,t+1} \quad (12)$$

where $\varepsilon_{3,t+1} \sim \text{i. i. d.}, N(0, \sigma_{\varepsilon_3}^2)$, and,

$$\alpha_3 = -\alpha_1, \beta_3 = -\beta_1, \varepsilon_{3,t+1} = -\varepsilon_{1,t+1}$$

Second, $\beta_1 > \beta_2$ is merely $\beta_1 - \beta_2 > 0$. To straightforwardly handle this, another relevant simple univariate regression (13), which is (1) deducting (2), is introduced below.

$$f_t - 2s_{t+1} + s_t = \alpha_4 + \beta_4(f_t - s_t) + \varepsilon_{4,t+1} \quad (13)$$

where $\varepsilon_{4,t+1} \sim \text{i. i. d.}, N(0, \sigma_{\varepsilon_4}^2)$, and,

$$\alpha_4 = \alpha_1 - \alpha_2, \beta_4 = \beta_1 - \beta_2, \varepsilon_{4,t+1} = \varepsilon_{1,t+1} - \varepsilon_{2,t+1}$$

Therefore, the estimated coefficient, $\beta_4$, can be expressed as (14) below ((8) deducting (9)). It is obvious that $\beta_4 > 0$ means $\sigma^2(rp_t) > \sigma^2(\Delta s_{t+1}^e)$.

$$\beta_4 = \beta_1 - \beta_2 = \frac{\sigma^2(rp_t) - \sigma^2(\Delta s_{t+1}^e)}{\sigma^2(rp_t + \Delta s_{t+1}^e)} \quad (14)$$

Consequently, $rp_t$ is existent and characterized time variant in $f_t - s_t$ if the following alternative hypotheses are statistically not rejected collectively.

$$H_1: \beta_3 \neq 0 \quad \text{(drawn from (12)), directly corresponding to (10)} \quad (15)$$
$$H_1: \beta_4 > 0 \quad \text{(drawn from (13)), directly corresponding to (11)} \quad (16)$$

Subsequent to defining unobservable foreign exchange risk premia and accepting those particular alternative hypotheses leading to the existence of time-evolved foreign exchange risk premia, the corresponding state-space setting may be further investigated.

## 3.2 State-space modeling

Once foreign exchange risk premia have been shown how they appear with unobservability (refer to (3)) and time variability (refer to (10) and (11)), the state-space representation may be developed along with the corresponding Kalman-filter algorithm and log-likelihood function as follows.

The main structure of a state-space model is basically composed of observation (or measurement or output or signal) and state (or transition) equations (i.e. (17) and (18), respectively, below). It may also be called a signal-extraction approach as a signal extracting its transition from a noisy environment. In examining unobservable foreign exchange risk premia, $rp_t$, they, as a state variable, are extracted from a noise or rational forecast errors, $re_{t+1}$, by a signal named forward forecast errors, $fe_t$ (Wolff, 1987).

A state variable with its time-variation property could basically follow $ARIMA(p, d, q)$. The process turns out to be $ARMA(p, q)$ if the state variable is stationary at level, i.e. integrated of order zero or $I(0)$. To consider the stationarity of the state or unobservable variable, it can be somewhat inferred by that of the observable variable. For the case of foreign-exchange-risk-premium analysis in specific, recall (5), $fe_t = rp_t + re_{t+1}$, which will be the observation equation. It can be seen that the behavior of $rp_t$ directly arises from that of $fe_t$ since $re_{t+1}$ is white noise (refer to (4)) (Cheung, 1993; Wolff, 1987). So, before simply proceeding an $ARMA(p, q)$ signal extraction, it is to have the state variable supposedly stationary in the beginning, that is primarily certifying the stationarity of the observable variable. In studying foreign exchange risk premia, $rp_t$, as the state or unobservable variable, the conforming observable variable is forward errors, $fe_t$. Accordingly, at the outset of $ARMA(p, q)$-based modeling, it follows that $\Delta(fe_t, d) \sim I(0)$, i.e. operate any differences on $fe_t$ to ascertain its stationarity. Nonetheless, fortunately, the time series of forward errors is commonly stationary without any differences, i.e. $\{fe_t\} \sim I(0)$ (Cheung, 1993; Wolff, 1987). Henceforth, it can be taken for granted that $\{rp_t\} \sim I(0)$; then, $\{rp_t\} \sim ARIMA(p, d = 0, q) = ARMA(p, q)$.

Although the time series $\{rp_t\}$ analytically and empirically comes up with considerably low $p$ or $q$ in $ARMA(p, q)$ in state-space researching, such as $AR(1)$, $MA(1)$, etc. (Cheung, 1993; Wolff, 1987), the generalization is still taken into consideration as to be accessible for any diagnostic case. To generalize a state-space analysis, it is thereby to display the state equation in such a form of $ARMA(p, q)$. Additionally, despite the Gaussian white-noise assumptions for the disturbances, the generalization may also include the fact that there possibly exists a co-movement between the residuals of the observation equation and those of the state equation, i.e. the residual covariance matrix can be either zero or non-zero, i.e. either $C = 0$ or $C \neq 0$ ($C = 0$ simplified by Wolff (1987) whereas $C \neq 0$ generalized by Cheung (1993)).[3] This is remarked because many cases purely assume $C = 0$ by theory, meant for simplicity. Applying to foreign exchange risk premia, such state-space generalization with $\{rp_t\} \sim ARMA(p, q)$ (in (18)) and $\sigma(\{re_{t+1}\}, \{a_t\}) = C$ (disturbance co-variation between (17) and (18)), becomes:

$$fe_t = rp_t + re_{t+1} \tag{17}$$

[4]

$$rp_t = \sum_{i=1}^{p} \phi_i (rp_{t-i}) + a_t + \sum_{j=1}^{q} \theta_j a_{t-j} \tag{18}$$

where $\{re_{t+1}\} \sim \text{i. i. d.}, N(0, R)$,

$\{a_t\}$ is risk-premium errors, $\{a_t\} \sim \text{i. i. d.}, N(0, Q)$,

$\sigma(\{re_{t+1}\}, \{a_t\}) = C$, and,

$R$, $Q$, and $C$ are Gaussian white-noise variance-covariance matrices[5]

The state-space system for the risk premium analysis above also requires the definitions in a forward-spot expression (i.e. as a function of $f_t$, $s_{t+1}^e$, or $s_{t+1}$) of $rp_t$, $re_{t+1}$, and $fe_t$ in which they are previously presented as (3), (4), and (5), respectively.

Following the state-space generalization, rewrite the transition equation (18) into a full vector format as (19) below (Cheung, 1993; Hamilton, 1994; Pelagatti, 2015).

$$\underset{(m \times 1)}{RP_t \equiv \begin{bmatrix} rp_{t,1} \\ \vdots \\ \vdots \\ rp_{t,m} \end{bmatrix}} = \underset{(m \times m)}{\begin{bmatrix} \phi_1 & | \ddots & & \iddots \\ \vdots & | & I_{m-1} & \\ \phi_{m-1} & | \underline{\iddots} & _ & \underline{\ddots} \\ \phi_m & 0 & \cdots & 0 \end{bmatrix}} \underset{(m \times 1)}{\begin{bmatrix} rp_{t-1,1} \\ \vdots \\ \vdots \\ rp_{t-1,m} \end{bmatrix}} + \underset{(m \times 1)}{\begin{bmatrix} 1 \\ \theta_1 \\ \vdots \\ \theta_{m-1} \end{bmatrix}} a_t \tag{19}$$

where $m = \max\{p, q + 1\}, \phi_i = 0$ if $i > p, \theta_j = 0$ if $j > q$

Next, in accordance with (19), re-express the forward-exchange-risk-premium state-space system into a vector set-up as demonstrated below, from (17) to (20) and from (18) to (21) in particular.

$$fe_t = Z(RP_t) + re_{t+1} \tag{20}$$

$$RP_t = \Phi(RP_{t-1}) + \Theta a_t \tag{21}$$

[3] 0 whenever in a matrix context, especially in state-space modeling, denotes a zero matrix.

[4] also recall (5)

[5] To repeat, $C = 0$ supposedly but $C \neq 0$ for relaxation or generalization.

where $\begin{pmatrix} re_{t+1} \\ a_t \end{pmatrix} \sim \text{i.i.d.}, N\left[\begin{pmatrix} 0 \\ 0 \end{pmatrix}, \begin{pmatrix} R & C \\ C & Q \end{pmatrix}\right]$, (22)

$Z = [1,0, \cdots ,0]$ with size $1 \times m$, and,

$\Phi$ and $\Theta$ already implicitly defined in (19) ((21) identically reduced from (19))

## Kalman-filter algorithm

A Kalman-filter algorithm is an algorithm useful for computing the mean and covariance matrices of the unobserved components in a state-space model throughout a procedure of period-by-period recursive updating (Hamilton, 1994; Pelagatti, 2015).

Provided the constructed $ARMA(p,q)$ state-space system of (20), (21), and (22) for studying foreign exchange risk premia that are unobserved and time dependent, the corresponding Kalman-filtering recursions are as per the equations of (23), (24), (25), (26), and (27) below (also see **Appendix A: Kalman-filter derivation**) (Cheung, 1993). To clarify, those are the equations defining the conditional mean, $E_t(\cdot)$, and conditional variance, $V_t(\cdot)$, of foreign exchange risk premia recursively over time.

$$E_t(RP_t) = E_{t-1}(RP_t) + K_t[fe_t - ZE_{t-1}(RP_t)] \quad (23)$$

$$V_t(RP_t) = V_{t-1}(RP_t) - K_t Z V_{t-1}(RP_t) \quad (24)$$

$$K_t = V_{t-1}(RP_t) Z^T [Z V_{t-1}(RP_t) Z^T + R]^{-1} \quad (25)$$

$$\begin{aligned} E_t(RP_{t+1}) = \Phi E_t(RP_t) + \\ + \Theta C[Z V_{t-1}(RP_t) Z^T + R]^{-1}[fe_t - ZE_{t-1}(RP_t)] \end{aligned} \quad (26)$$

$$\begin{aligned} V_t(RP_{t+1}) = \Phi V_t(RP_t)\Phi^T + \Theta Q \Theta^T - \Theta C[Z V_{t-1}(RP_t) Z^T + R]^{-1}\Theta^T C^T + \\ - \Phi K_t C^T \Theta^T - \Theta C K_t^T \Phi^T \end{aligned} \quad (27)$$

The Kalman-filter algorithm becomes standardized if $C = 0$ (Cheung, 1993; Wolff, 1987). Nevertheless, this simplification might cause misspecification in foreign-exchange-risk-premium modeling; thus, it is rather to let $C \neq 0$ possible, i.e. the forward errors, $fe_t$, may contain some meaningful information regarding the risk-premium residual, $a_t$. If $fe_t$ is statistically related to $a_t$, $C \neq 0$ will help explain $rp_t$ (Cheung, 1993). If not, the algorithmically computed $C$ will just show up approaching 0.

## Log-likelihood function

Denote rational-premium forecast error differentials as $\epsilon_t$ and refer to (5), then:

$$\epsilon_t = (s_{t+1}^e - s_{t+1}) - (rp_t^e - rp_t) = fe_t - rp_t^e \quad (28)$$

In consequence of taking account for the rational-premium forecaslt error differentials, $\epsilon_t$, with its consequential full matrix, $\xi_t$, the log-likelihood function ($L$) evaluated by the Kalman filter algorithm of the foreign-exchange-risk-premium state space ((23)-(27)) is formulated as (Cheung, 1993):

$$L = -\frac{T}{2}\ln 2\pi - \frac{1}{2}\sum_{t=1}^{T} \ln[V_{t-1}(RP_t) + R] - \frac{1}{2}\sum_{t=1}^{T} \frac{\xi_t^2}{V_{t-1}(RP_t)+R} \quad (29)$$

The Kalman filtering, i.e. (23)-(27), together with the resultant log-likelihood function, (29), will thereafter estimate the state-space matrix parameters, inclusive of $\Phi, \Theta, R, Q$, and $C$, present in the signal-extraction system of (20), (21), and (22) via a maximum-likelihood methodology (Cheung, 1993; Wolff, 1987).

## Relevant intuitions on risk premia

As the two main generalization settings in the state-space modeling are raised, which are $\{rp_t\}\sim ARMA(p,q)$ and possible $C \neq 0$, they may be intuitively explained more below.

Hypothetically, closer to the foreign exchange market efficiency, nearer to be white noise the risk premia, $rp_t$, are. That is the values of $p$ and $q$ in $\{rp_t\}\sim ARMA(p,q)$ are smaller. Virtually, these should get along with an exchange rate with longer data series and a higher degree of market determination. Contrariwise, higher $p$ or $q$ in $\{rp_t\}\sim ARMA(p,q)$ states that $\{rp_t\}$ is farer from being white noise. This points toward less market efficiency of the underlying currency.

Then, take a look at some perception behind the association between forward errors, $fe_t$, and risk-premium residuals, $a_t$ (recall the covariance $\sigma(\{re_{t+1}\},\{a_t\}) = C$). Recall the state equation (18), $\{rp_t\}\sim ARMA(p,q)$: $rp_t = \sum_{i=1}^{p} \phi_i(rp_{t-i}) + a_t + \sum_{j=1}^{q} \theta_j a_{t-j}$. The risk premia residual term or idiosyncratic part is $a_t$. So, the rest, which is $\sum_{i=1}^{p} \phi_i(rp_{t-i}) + \sum_{j=1}^{q} \theta_j a_{t-j}$, is systematic (Wolff, 1987). Symbolize the systemic risk premia as $rp_t^s$.

$$rp_t^s = \sum_{i=1}^{p} \phi_i(rp_{t-i}) + \sum_{j=1}^{q} \theta_j a_{t-j} \quad (30)$$

Recall (5) $fe_t = rp_t + re_{t+1}$; then:

$$fe_t - rp_t^s = re_{t+1} + a_t \quad (31)$$

(31) is that forward prediction errors excluding the systemic risk premia amounts to the combination of rational prediction errors and idiosyncratic risk premia (sometimes known as innovations or shocks). Thence, $fe_t - rp_t^s$ should be white noise if $re_{t+1}$ and $a_t$ are appropriately independent and uncorrelated signifying that $C = 0$ and vice versa in connection with $C \neq 0$.

## 4. Data description and statistics

Let the US be the domestic country. The monthly data set employed and analyzed contains foreign exchange spot and one-month forward rates of one well-settled currency pair and the other pair, which is more newly settled in the global economy. They are USD/GBP and USD/HKD, respectively (i.e. domestic/foreign). USD/GBP is from January 1979 to March 2016 and USD/HKD is from January 1998 to March 2016. All the monthly forwards data series available up to the moment are taken into account. Statistical Interactive Database of Bank of England[6] is the source for the former and Market Data & Statistics of Hong Kong Monetary Authority[7] is for the latter. The econometric software package used in this paper is EViews.

This paper is to study how foreign exchange risk premia behave temporally. If risk premia stem from a fixed currency, it will then instead study how the premia are someway manipulated. Hence, it is also essential that the analytical currencies should be well mechanized by market demands and supplies as to rationally reflect the foreign exchange equilibrium and market efficiency in a longer run as theorized. For USD/GBP, it has been properly floating since late 1970's after the end of the Bretton Woods system. For the USD/HKD forward rates, though the data is obtainable since January 1996, it is rather to be explored since January 1998 instead. This is due to the fact that the underlying currency became fundamentally floating since mid to late 1997 after strictly managed anchoring to USD until the 1997 Asian currency crisis.

Accordingly, the description and statistics of the data, which is particularly the inferring observed time sequence namely forward errors, $\{fe_t\}$, are in **Table 1** below in which such $\{fe_t\}$ is up to February 2016 since $fe_t = f_t - s_{t+1}$ (i.e. the last observation: $fe_{Feb2016} = f_{Feb2016} - s_{Mar2016}$). The descriptions and statistics include the first, second, third, and fourth moments (mean, s.d., skewness, and excess kurtosis, respectively), and Jarque–Bera (JB) and Augmented Dickey-Fuller (ADF)[8] tests of $\{fe_t\}$ (the null hypotheses are normality and non-stationarity, respectively).

[6] http://www.bankofengland.co.uk/boeapps/iadb/index.asp?first=yes&SectionRequired=I&HideNums=-1&ExtraInfo=true&Travel=NIx
[7] http://www.hkma.gov.hk/eng/market-data-and-statistics/monthly-statistical-bulletin/table.shtml#section6
[8] ADF test with intercept and trend

| | USD/GBP | USD/HKD |
|---|---|---|
| observations | 446 | 218 |
| mean | -0.000573 | 7.04E-05 |
| s.d. | 0.029831 | 0.001633 |
| skewness | 0.184238 | -1.397168 |
| excess kurtosis | 1.699602 | 9.982169 |
| JB stat | 56.20380 | 1083.474 |
| JB p-value | 0.000000 | 0.000000 |
| ADF t-stat | -19.05458 | -12.23439 |
| ADF p-value | 0.000000 | 0.000000 |

**Table 1**: description and statistics of $\{fe_t\}$

Loot at **Table 1**. For USD/GBP, $\{fe_t\}$ is slightly positively skewed and also to some extent leptokurtic. In spite of non-normality, it is reliably stationary at level. For USD/HKD, $\{fe_t\}$ is somewhat negatively skewed and noticeably leptokurtic. Even though it is not statistically normally distributed, its level is properly stationary. Importantly, $\{fe_t\}\sim I(0)$ for both currency pairs thereby implies that $\{rp_t\}\sim I(0)$ and $\{rp_t\}\sim ARIMA(p, d=0, q) = ARMA(p, q)$ for both of them as anticipated in generalizing a state-space model as earlier shown (refer to **3.2 State-space modeling**).

# 5. Empirics

The empirics of foreign-exchange-risk-premium analyses in this paper comprise empirical methodology, traditional-based results, and state-space results as described below.

## 5.1 Empirical methodology

In this context, the foreign-exchange-risk-premium analyses mainly begin with traditional univariate regressions and then state-space implementations. The state-space procedure subsequent to obtaining and evaluating $\beta_3$ and $\beta_4$ from the simple univariate regressions of (12) and (13) could be done along these lines and paragraphs.

After ensuring $\{rp_t\}\sim I(0)$ and $\{rp_t\}\sim ARMA(p, q)$ such that the analytical state-space representation is in the generalized form of $ARMA(p, q)$ (refer to **4. Data description and statistics** by which $fe_t$ infers $rp_t$ as already explained in more detail in **3.2 State-space modeling**), then proceed the following Box-Jenkins (BJ) methodology for the $ARMA(p, q)$ (Gujarati and Porter, 2009).[9]

Initially, $p$ and $q$ are identified throughout observing the partial autocorrelation (PAC) and autocorrelation (AC) functions of the referenced observable series $\{fe_t\}$ under which those functions are displayed up to the twelfth lag owing to the data's monthly basis (see **Table 2**). To clarify, PAC hints at $p$ and AC at $q$.[10]

[9] The steps of the BJ methodology for $ARMA(p, q)$: identifying $p$ and $q$, estimating $ARMA(p, q)$, and diagnosing the white-noise process and stationarity of the residual.

| | USD/GBP | | | | USD/HKD | | | |
|---|---|---|---|---|---|---|---|---|
| lag | PAC | | AC | | PAC | | AC | |
| 1 | 0.097 | ** | 0.097 | ** | 0.207 | *** | 0.207 | *** |
| 2 | 0.042 | | 0.051 | | 0.034 | | 0.075 | |
| 3 | 0.050 | | 0.058 | | 0.014 | | 0.035 | |
| 4 | 0.005 | | 0.017 | | 0.052 | | 0.062 | |
| 5 | -0.003 | | 0.003 | | 0.099 | | 0.119 | * |
| 6 | -0.055 | | -0.050 | | 0.039 | | 0.085 | |
| 7 | -0.073 | | -0.080 | * | 0.051 | | 0.084 | |
| 8 | 0.029 | | 0.010 | | -0.038 | | -0.000 | |
| 9 | 0.025 | | 0.016 | | -0.038 | | -0.026 | |
| 10 | -0.044 | | -0.048 | | 0.030 | | 0.032 | |
| 11 | 0.066 | | 0.058 | | -0.012 | | 0.014 | |
| 12 | 0.006 | | 0.017 | | 0.064 | | 0.077 | |

**Table 2**: PAC and AC of $\{fe_t\}$ [11]
***, **, * indicate significance at the 1%, 5%, 10% levels, respectively
using standard error $1/\sqrt{T}$ [12]

From **Table 2**, by typically detecting up to 5% significance level, $ARMA(p = 1, q = 1)$ may be implied for both currency pairs.[13] Both time-series processes $AR$ and $MA$ are found to be with as low as lag order 1, which is remarkably small as hypothesized.

Based on the statistical significance of $p = 1$ and $q = 1$ suggested beforehand by the PAC and AC of the underlying $\{fe_t\}$, certain consequential $ARMA(p, q)$ processes for $\{fe_t\}$ are afterwards estimated via employing a trial and error method. The model selection depends on lowest Akaike information criterion (AIC), lowest Schwarz criterion (SC), or lowest Hannan-Quinn criterion (HQC)[14] alongside the diagnostic check of the resulting disturbance. Test whether the residual is a white-noise process via Ljung-Box (LB) Q statistic[15] up to a certain lag with the null hypothesis of independent distribution, whether serially uncorrelated via Breusch-Godfrey (BG) test[16] with the null hypothesis

[10] AC reflects $MA(q)$ and PAC reflects $AR(p)$.

[11] Representing PAC before AC unlike statistical packages in order to reflect $AR(p)$ then $MA(q)$ in $ARMA(p, q)$.

[12] significance values for PAC and AC = (standard units of standard errors)*$(1/\sqrt{T})$
For USD/GBP with $T = 446$, the significance values at the 1%, 5%, and 10% levels for PAC and AC are $\pm 0.12197$, $\pm 0.09281$, and $\pm 0.07789$, respectively.
For USD/HKD with $T = 218$, the significance values at the 1%, 5%, and 10% levels for PAC and AC are $\pm 0.17446$, $\pm 0.13275$, and $\pm 0.11140$, respectively.

[13] For all USD/HKD $fe_t$ available since January 1996 including its fixed-regime period (up to mid-late 1997), AC is significant at lags 5, 10, and 12 as well. The correlations significant at some higher lags might be implicit that such fixed USD/HKD to some extent represents market inefficiency.

[14] Despite slightly distinctive computations, lower AIC, SC, or HQC all sensibly and statistically indicates better model specification.

[15] LB Q stat shown up at several relevant lags in **Tables 3** and **4** is due to potential practical problem of lag selection (refer to http://www.eviews.com/help/helpintro.html#page/EViews%209%20Help/series.018.07.html).

[16] BG test up to lag 2, done in EViews 7 due to unavailability in EViews 9.5 Student Version Lite.

of non-existing serial correlations, and whether stationary via ADF test[17] with the null hypothesis of non-stationarity (see **Tables 3** and **4**).

| | | | $\boldsymbol{ARMA(1,1)}$ | $\boldsymbol{AR(1)}$ | $\boldsymbol{MA(1)}$ |
|---|---|---|---|---|---|
| model selection criteria | AIC | | -4.187524 | <u>-4.188971</u> | -4.188241 |
| | SC | | -4.159943 | <u>-4.170584</u> | -4.169854 |
| | HQC | | -4.176649 | <u>-4.181722</u> | -4.180991 |
| disturbance | LB Q p-value | up to lag 12 | 0.604 | 0.614 | 0.589 |
| | | up to lag 24 | 0.716 | 0.728 | 0.708 |
| | | up to lag 36 | 0.883 | 0.873 | 0.859 |
| | BG $\chi^2$ p-value | | 1.0000 | 0.4412 | 0.3189 |
| | ADF p-value | | 0.0000 | 0.0000 | 0.0000 |

**Table 3**: $ARMA(p, q)$ trial and error for the $\{fe_t\}$ of USD/GBP
(the underlined indicates selected by a certain model selection criterion)

| | | | $\boldsymbol{ARMA(1,1)}$ | $\boldsymbol{AR(1)}$ | $\boldsymbol{MA(1)}$ |
|---|---|---|---|---|---|
| model selection criteria | AIC | | -10.00885 | <u>-10.01494</u> | -10.00972 |
| | SC | | -9.962274 | <u>-9.983890</u> | -9.978674 |
| | HQC | | -9.990037 | <u>-10.00240</u> | -9.997183 |
| disturbance | LB Q p-value | up to lag 12 | 0.702 | 0.872 | 0.784 |
| | | up to lag 24 | 0.690 | 0.845 | 0.785 |
| | | up to lag 36 | 0.738 | 0.917 | 0.889 |
| | BG $\chi^2$ p-value | | 0.6156 | 0.4139 | 0.1935 |
| | ADF p-value | | 0.0000 | 0.0000 | 0.0000 |

**Table 4**: $ARMA(p, q)$ trial and error for the $\{fe_t\}$ of USD/HKD
(the underlined indicates selected by a certain model selection criterion)

**Tables 3** and **4** demonstrate that the selected model is $AR(1)$ for both USD/GBP and USD/HKD with lowest AIC, SC, and HQC altogether as well as the residuals appropriately being white noise, serial uncorrelated, and stationary.

## $\boldsymbol{ARMA(p, q)}$ for foreign exchange risk premia

Recall (5) $fe_t = rp_t + re_{t+1}$. The rational error, $re_{t+1}$, is white noise therefore following either $AR(0)$ or $MA(0)$.

$$\{re_{t+1}\} \sim \begin{cases} AR(0) \\ MA(0) \end{cases}$$

In consequence, either a pure $AR(p)$ or a pure $MA(q)$ for $\{fe_t\}$ determines the identical form of such either $AR(p)$ or $MA(q)$ for $\{rp_t\}$. Applicably, $AR(1)$ for $\{fe_t\}$ implies $AR(1)$ for $\{rp_t\}$ pertaining to both of two foreign exchanges.[18]

[17] ADF test with intercept and trend
[18] So, just in case, $MA(1)$ for $\{fe_t\}$ implies $MA(1)$ for $\{rp_t\}$, accordingly.

$$\{rp_t\} \sim AR(1) \tag{32}$$

Additionally, notwithstanding to go on with the initial $\{fe_t\} \sim ARMA(1,1)$ as evidently directed by the significant statistics of existing autocorrelation and partial autocorrelation functions without implementing such trial and error, the resultant process is still $\{rp_t\} \sim AR(1)$ not $\{rp_t\} \sim ARMA(1,1)$ as clarified below.

Express $\{fe_t\} \sim ARMA(1,1)$ with its corresponding white-noise $re_{t+1}$ (refer to (4)) as:

$$fe_t = \gamma_{fe}(fe_{t-1}) + re_{t+1} + \gamma_{re}(re_t)$$

And, again, recall (5) $fe_t = rp_t + re_{t+1}$. After substitutions and simplification, then:

$$rp_t + re_{t+1} = \gamma_{fe}(rp_{t-1}) + \gamma_{fe}(re_t) + re_{t+1} + \gamma_{re}(re_t)$$

$$rp_t = \gamma_{fe}\left(rp_{t-1}\right) + 2\gamma_{fe}(re_t) \tag{33}$$

(33) shows a form of $AR(1)$ with white-noise $2\gamma_{fe}(re_t)$ so that $\{rp_t\} \sim AR(1)$, which is consistent with (32) (also in agreement with Wolff (1987)).

After all, only (32) is to be further proceeded in the state-space analyses of foreign exchange risk premia.

## State-space specification

Thereafter, upon the identified process (32), the $ARMA(p,q)$-state-space formulation and estimation (with $fe_t$ as the observed and $rp_t$ as the unobserved; or, the system of (20)-(22) through the Kalman-filter algorithm[19] of (23)-(27), specifically) are afterwards carried out via EViews[20] (with manual and automatic state-space specifications helped explain by Van den Bossche (2011) and EViews website[21]). Mainly, there are three parts compatible with the fundamental state-space structure: signal or observation (@signal), state (@state), and residuals (@evar).

The state-space system of (20)-(22) for $\{rp_t\} \sim AR(1)$ (refer to (32)) with possible $C \neq 0$ can be formulated in EViews as shown in **Figure 1** below.

---

[19] EViews automatically computes the Kalman filter recursively based on a certain state-space specification.

[20] The object in EViews for state-space specifications is "SSpace", i.e. Object/New Object.../Sspace.

[21] http://www.eviews.com/help/helpintro.html#page/EViews%25209%2520Help%2Fsspace.053.3.html%23

```
@signal fe = sv1 + [ename = e1]

@state sv1 = c(4)*sv1(-1) + [ename = e2]

@evar var(e1) = exp(c(1))
@evar var(e2) = exp(c(2))
@evar cov(e1, e2) = c(3)
```

**Figure 1**: $AR(1)$-state-space formulation with possible $C \neq 0$

If $C = 0$ simplified by the theoretical assumption, '@evar cov(e1, e2) = 0' is defined or that '@evar cov(e1, e2)' disappears (may also see EViews formulations for $ARMA(1,1)$ and $MA(1)$ in **Appendix B: Additional state-space formulations**, which might be useful for other financial time series).

For state-space estimations through the Kalman-filter algorithm, the BFGS optimization method and the Marquardt step method are employed by default in EViews 9.5 Student Version Lite to optimize the log likelihood function.

## 5.2 Traditional-based results

Recall the alternative hypotheses (15) and (16) such that $H_1: \beta_3 \neq 0$ suggests $s_{t+1} \neq f_t$ (also refer to (10)) and $H_1: \beta_4 > 0$ does $\sigma^2(rp_t) > \sigma^2(\Delta s_{t+1}^e)$ (also refer to (11)). These characterize the presence and time evolvement of foreign exchange risk premia, $rp_t$, as previously explained (in **Time-varying foreign exchange risk premia**). The univariate regression outcomes of time-evolved $rp_t$ for USD/GBP and USD/HKD are then shown in **Table 5** below.

| | $H_1$ | statistics | USD/GBP | USD/HKD |
|---|---|---|---|---|
| | | $\beta_3$ | -3.119972 | -1.021640 |
| | | standard error | 0.713626 | 0.089838 |
| $s_{t+1} \neq f_t$ | $\beta_3 \neq 0$ | t-stat[22] | -4.371999 | -11.372000 |
| | | two-tail p-value | 0.000015 | 0.000000 |
| | | disturbance ADF[23] p-value | 0.000000 | 0.000000 |
| | | $\beta_4$ | 5.239943 | 1.043281 |
| | | standard error | 1.427252 | 0.179676 |
| $\sigma^2(rp_t) > \sigma^2(\Delta s_{t+1}^e)$ | $\beta_4 > 0$ | t-stat | 3.671351 | 5.806441 |
| | | one-tail p-value[24] | 0.000135 | 0.000000 |
| | | disturbance ADF[20] p-value | 0.000000 | 0.000000 |

**Table 5**: univariate regression results as regards $rp_t$ time variation

Both alternative hypotheses are statistically accepted with suitably stationary residuals for both USD/GBP and USD/HKD. Henceforward, the expressions $s_{t+1} \neq f_t$ and

[22] The sign is opposite to the t-stat for $\beta_1$ as aforementioned.
[23] ADF test with intercept and trend
[24] The originally obtained two-tail p-value for USD/GBP is 0.000270643 and USD/HKD is 2.26e-8.

$\sigma^2(rp_t) > \sigma^2(\Delta s^e_{t+1})$ are meaningful. Thus, it can be to a certain extent concluded that foreign exchange risk premia concerning both studied currency pairs appear significantly and varyingly through time.

## 5.3 State-space results

The foreign exchange risk premia of both USD/GBP and USD/HKD are found reliably existent and temporal. Hereafter, it can be worthwhile to utilize an $ARMA(p, q)$-state-space approach to further examine unobserved behaviors within such discovered time-series property. The process solely projected is (32), $\{rp_t\} \sim AR(1)$.

The estimated $AR(1)$-state-space model with fitted parameters $(\hat{\cdot})$ (statistical results in **Tables 6** and **8**) looks like:

$$fe_t = \widehat{rp}_t + \widehat{re}_{t+1}$$
$$\widehat{rp}_t = \hat{\phi}_1(\widehat{rp}_{t-1}) + \hat{a}_t$$

with $\hat{R} = \sigma^2(\widehat{re}_{t+1}), \hat{Q} = \sigma^2(\hat{a}_t)$, and $\hat{C} = \sigma(\widehat{re}_{t+1}, \hat{a}_t)$

The sequence $\{\widehat{rp}_t\}$, so-called the predicted state series, is directly generated by the model (see **Figure 2** for USD/GBP and **Figure 3** for USD/HKD later). Then, $\widehat{re}_{t+1}$ and $\hat{a}_t$ are simply computed as follows.

$$\widehat{re}_{t+1} = \widehat{rp}_t - fe_t$$
$$\hat{a}_t = \widehat{rp}_t - \hat{\phi}_1(\widehat{rp}_{t-1})$$

Subsequently, the following forward prediction errors removing systematic risk premia (recall (31), $fe_t - rp^s_t = re_{t+1} + a_t$) can be calculated by summing up the fitted rational expectation errors, $\widehat{re}_{t+1}$, and the fitted risk-premium residuals, $\hat{a}_t$.

$$fe_t - \widehat{rp}^s_t = \widehat{re}_{t+1} + \hat{a}_t \qquad (34)$$

Such disturbance combination (34) is further analyzed later as also represented in **Tables 7** and **9**.

Numerically and visually, the empirical results, inclusive of the projected parameters ($\hat{\Phi}, \hat{R}, \hat{Q}$, and $\hat{C}$), maximized log-likelihood ($L$), model criteria (AIC, SC, and HQC), disturbance combination ($\widehat{re}_{t+1} + \hat{a}_t$), and predicted risk premia ($\widehat{rp}_t$) are presented below currency by currency, USD/GBP then USD/HKD. Broadly, $C \neq 0$ relaxation does not make a significant difference ending up with critically analogous results to the preliminary $C = 0$ scheme.

## USD/GBP

| | $\boldsymbol{C=0}$ | $\boldsymbol{C\neq 0}$ |
|---|---|---|
| $\widehat{\Phi}=\hat{\phi}_1$ | 0.550056 | 0.550056 |
| p-value on $\phi_1$ | 0.047635 | 0.048170 |
| $\hat{R}=\sigma^2(\widehat{re}_{t+1})$ | 0.000727 | 4.953e-12 |
| $\hat{Q}=\sigma^2(\hat{a}_t)$ | 0.000112 | 0.000619 |
| $\hat{C}=\sigma(\widehat{re}_{t+1},\hat{a}_t)$ | 0 | -0.000400 |
| $\sigma^2(\widehat{rp}_t)$ | 1.109e-5 | 5.295e-6 |
| $L$ | 936.817739 | 936.817730 |
| AIC | -4.187523 | -4.183039 |
| SC | -4.159943 | -4.146265 |
| HQC | -4.176649 | -4.168540 |

**Table 6**: $AR(1)$-state-space results for USD/GBP

**Table 6** belonging to USD/GBP shows that relaxing $C\neq 0$ does not improve the model with the theoretic supposition of $C=0$. The $C=0$ set-up looks slightly superior in all terms of p-value on $\phi_1$ (lower better), $L$ (higher better), and AIC, SC, and HQC (criteria lower better) although by considerably marginal quantities. Yet, the negative direction of $C$ suggests meaningfully that forward and premium errors ($re_{t+1}$ and $a_t$) counter-move each other even if insignificantly. Moreover, the $AR(1)$ coefficient on risk premia is statistically significant at 5% level for the two cases. As $fe_t$ comprise $rp_t$ and $re_{t+1}$, $\sigma^2(\widehat{rp}_t)>\sigma^2(\widehat{re}_{t+1})$ obviously for $C\neq 0$ but quite not for $C=0$ say that risk-premium variance is more responsible than rational-error variance for the variability of forward prediction errors only in some measure.

| | $\boldsymbol{C=0}$ | | | | $\boldsymbol{C\neq 0}$ | | | |
|---|---|---|---|---|---|---|---|---|
| lag | PAC | | AC | | PAC | | AC | |
| 1 | 0.088 | * | 0.088 | * | 0.088 | * | 0.088 | * |
| 2 | -0.011 | | -0.003 | | -0.011 | | -0.003 | |
| 3 | 0.033 | | 0.031 | | 0.033 | | 0.031 | |
| 4 | -0.002 | | 0.004 | | -0.002 | | 0.004 | |
| 5 | -0.005 | | -0.005 | | -0.005 | | -0.005 | |
| 6 | -0.058 | | -0.057 | | -0.058 | | -0.057 | |
| 7 | -0.073 | | -0.082 | * | -0.073 | | -0.082 | * |
| 8 | 0.029 | | 0.015 | | 0.029 | | 0.015 | |
| 9 | 0.014 | | 0.015 | | 0.014 | | 0.015 | |
| 10 | -0.045 | | -0.047 | | -0.045 | | -0.047 | |
| 11 | 0.068 | | 0.060 | | 0.068 | | 0.060 | |
| 12 | 0.008 | | 0.024 | | 0.008 | | 0.024 | |

| **LB Q p-value** | | |
|---|---|---|
| up to lag | $C=0$ | $C\neq 0$ |
| 12 | 0.472 | 0.472 |
| 24 | 0.577 | 0.577 |
| 36 | 0.810 | 0.810 |

**Table 7**: PAC, AC, and LB Q p-values of $\{fe_t-\widehat{rp}_t^s=\hat{a}_t+\widehat{re}_{t+1}\}$ for USD/GBP [25]
***, **, * indicate significance at the 1%, 5%, 10% levels, respectively
using standard error $1/\sqrt{T}$ [26]

[25] Representing PAC before AC unlike statistical packages in order to reflect $AR(p)$ then $MA(q)$ in $ARMA(p,q)$.
[26] With $T=445$ after adjustment, the significance values at the 1%, 5%, and 10% levels for PAC and AC are $\pm 0.12211$, $\pm 0.09291$, and $\pm 0.07797$, respectively.

The insignificance of PAC and AC (considered up to 5% significance level as usual) and LB Q statistic interpreted from **Table 7** implies that the combination of the two disturbance terms ($re_{t+1}$ and $a_t$) drawn from the $AR(1)$-state-space model is white noise. This fairly affirms that the model implemented for USD/GBP succeeds in capturing the essence of risk-premium time evolvement. In other words, getting rid of the systematic part of risk premia can seize serial correlations in forward errors. After all, these are in line with $C = 0$ in general.

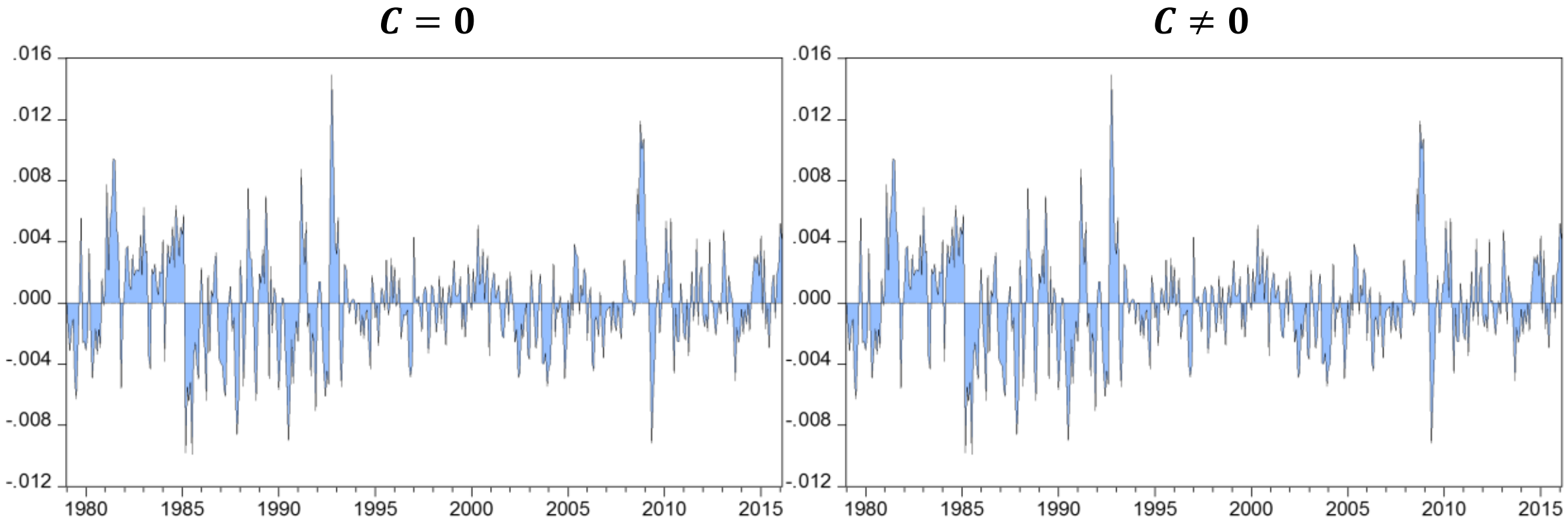

**Figure 2**: $\{\widehat{rp}_t\}$ of USD/GBP with $C = 0$ on the left and $C \neq 0$ on the right

**Figure 2** above supports that foreign exchange risk premia substantially fluctuate over time by way of fairly clumping pattern are also significantly stationary at level.[27] Overall, the frequencies of premia ($rp_t > 0$ i.e. $f_t > s^e_{t+1}$) and discounts ($rp_t < 0$ i.e. $f_t < s^e_{t+1}$) seem to occur evenly with some occasionally noticeable clustering.

## USD/HKD

| | $\boldsymbol{C = 0}$ | $\boldsymbol{C \neq 0}$ |
|---|---|---|
| $\hat{\Phi} = \hat{\phi}_1$ | 0.548398 | 0.548398 |
| p-value on $\phi_1$ | 0.011632 | NA[28] |
| $R = \sigma^2(\widehat{re}_{t+1})$ | 1.667e-6 | 1.667e-6 |
| $Q = \sigma^2(\hat{a}_t)$ | 7.218e-7 | 7.218e-7 |
| $C = \sigma(\widehat{re}_{t+1}, \hat{a}_t)$ | 0 | -0.000059 |
| $\sigma^2(\widehat{rp}_t)$ | 1.325e-7 | 1.325e-7 |
| $L$ | 1093.885274 | 1093.885274 |
| AIC | -10.008121 | -9.998947 |
| SC | -9.961546 | -9.936847 |
| HQC | -9.989309 | -9.973864 |

**Table 8**: $AR(1)$-state-space results for USD/HKD

[27] The p-value of the unit-root ADF test with intercept and trend is as low as 2.41e-21.

[28] EViews fails to iterate and produce any higher $L$ after letting $C \neq 0$ coming up with some inconsistent results. Consequently, such p-value is not available (NA).

Similar to the instance of USD/GBP, $C \neq 0$ generalization for USD/HKD is less favorable by means of model selection criteria as yielding higher AIC, SC, and HQ entirely than imposing $C = 0$ (see **Table 8**). Again, regardless of insignificance, the directions of $re_{t+1}$ and $a_t$ are opposite. Furthermore, the parameter of the $\{rp_t\} \sim AR(1)$ process is reliable at 5% and even nearby 1% significance level. For $\sigma^2(\widehat{rp}_t) < \sigma^2(\widehat{re}_{t+1})$ (pertaining to both $C = 0$ and $C \neq 0$) like USD/GBP with $C = 0$, risk-premium is rather less prevalent than rational-error variation in forward-error variation.

| | $C = 0$ | | | | $C \neq 0$ | | | |
|---|---|---|---|---|---|---|---|---|
| lag | PAC | | AC | | PAC | | AC | |
| 1 | 0.184 | *** | 0.184 | *** | 0.184 | *** | 0.184 | *** |
| 2 | -0.085 | | -0.049 | | -0.085 | | -0.049 | |
| 3 | -0.018 | | -0.043 | | -0.018 | | -0.043 | |
| 4 | 0.012 | | 0.004 | | 0.012 | | 0.004 | |
| 5 | 0.077 | | 0.082 | | 0.077 | | 0.082 | |
| 6 | 0.033 | | 0.063 | | 0.033 | | 0.063 | |
| 7 | 0.001 | | 0.009 | | 0.001 | | 0.009 | |
| 8 | -0.023 | | -0.031 | | -0.023 | | -0.031 | |
| 9 | -0.035 | | -0.047 | | -0.035 | | -0.047 | |
| 10 | 0.028 | | 0.020 | | 0.028 | | 0.020 | |
| 11 | -0.010 | | 0.013 | | -0.010 | | 0.013 | |
| 12 | 0.073 | | 0.072 | | 0.073 | | 0.072 | |

| LB Q p-value | | |
|---|---|---|
| up to lag | $C = 0$ | $C \neq 0$ |
| 12 | 0.381 | 0.381 |
| 24 | 0.515 | 0.515 |
| 36 | 0.542 | 0.542 |

**Table 9**: PAC, AC, and LB Q p-values of $\{fe_t - \widehat{rp}_t^s = \hat{a}_t + \widehat{re}_{t+1}\}$ for USD/HKD [29]
***, **, * indicate significance at the 1%, 5%, 10% levels, respectively
using standard error $1/\sqrt{T}$ [30]

Nonetheless, the $AR(1)$-state-space approach is not quite completely successful in coping with the prominence in the time-series property of USD/HKD risk premia as PAC and AC are insistently significant at lag order 1 (see **Table 9**). Despite this disappointment, the residual combination from the model is considered white-noise according to the computed LB Q statistic p-values at several related lags, which is in proportion to $C = 0$.

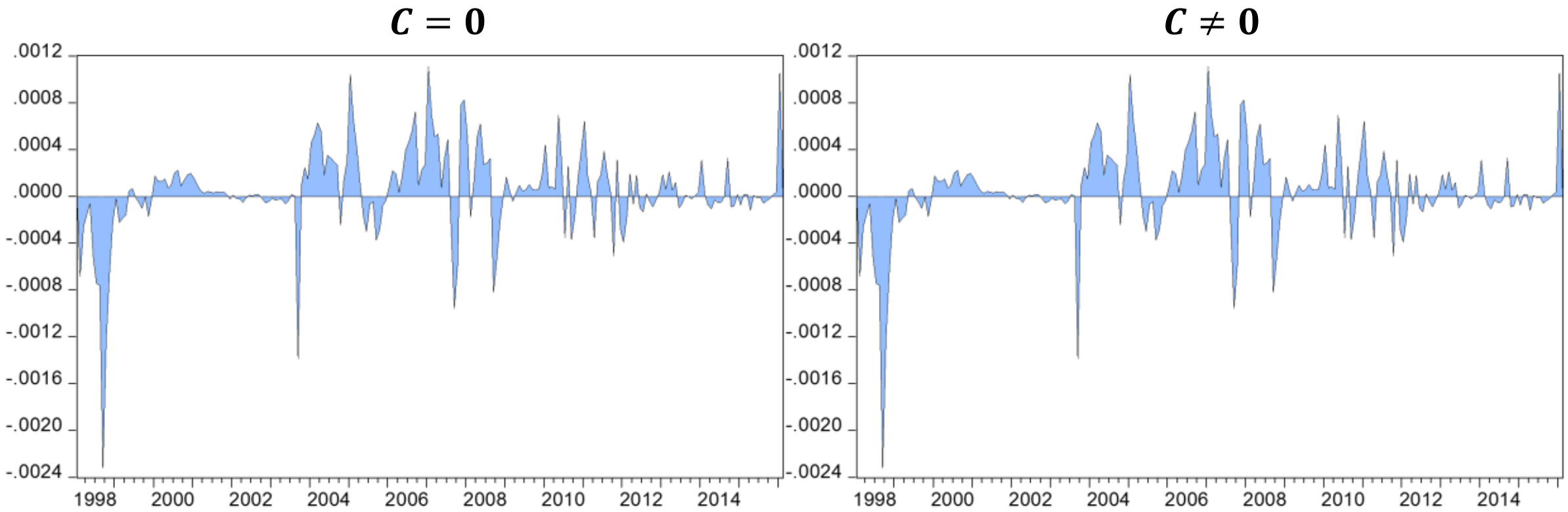

**Figure 3**: $\{\widehat{rp}_t\}$ of USD/HKD with $C = 0$ on the left and $C \neq 0$ on the right

[29] Representing PAC before AC unlike statistical packages in order to reflect $AR(p)$ then $MA(q)$ in $ARMA(p, q)$.
[30] with $T = 217$ after adjustment, the significance values at the 1%, 5%, and 10% levels for PAC and AC are $\pm 0.17486$, $\pm 0.13305$, and $\pm 0.11166$, respectively.

In the face of significant risk-premium time variation found by the univariate regressions, **Figure 3** however demonstrates that the risk premia of USD/HKD, which are found statistically stationary at level[31], are less volatile but more clustering than those of USD/GBP. The occurrences of premia ($rp_t > 0$ i.e. $f_t > s^e_{t+1}$) and discounts ($rp_t < 0$ i.e. $f_t < s^e_{t+1}$) are not quite balanced with more on the former. There exist comparably greater forward discounts earlier from 1998 to early 1999 after the 1997 Asian financial crisis. This could be owing to excessive depreciation expectations instantly following the currency-based economic shock. Conversely in years afterwards, USD/HKD became rather premia offsetting the previous undue risk discounts. Encouragingly, the currency has been more settled through time, especially via payments and transactions in Pacific regions.

## 6. Conclusion

Overwhelmingly, the traditional-grounded univariate regressions draw out significantly the existence and time-variant characteristic of foreign exchange risk premia. In stepping further, the state-space structuring is to a certain extent effective in investigating the unobservability and fundamentals of risk-premium time variability. In particular, the estimated parameters on the lagged projected time-series are statistically significant and the disturbance pooled from the signal and state equations is reliably white noise. Regarding the association between those signal and state residuals, namely forward and premium errors, their co-movement is negative while the covariance is recommended by the experiments to be zero. In addition, over time, stationary risk premia are varying through some clustering. Nonetheless, such risk-premium variability is not justly considered primary in forward-error fluctuations.

[31] The p-value of the unit-root ADF test with intercept and trend is as low as 5.28e-11.

# Appendices

## Appendix A: Kalman-filter derivation

The proof of the Kalman-filter algorithm with possible $C \neq 0$, (23)-(27), following the state-space system of (20)-(22), is according to Cheung (1993) as follows.

Recall the foreign-exchange-risk-premium state-space system (20)-(22).

$$fe_t = Z(RP_t) + re_{t+1}$$
$$RP_t = \Phi(RP_{t-1}) + \Theta a_t$$

where $\begin{pmatrix} re_{t+1} \\ a_t \end{pmatrix} \sim \text{i.i.d.}, N\left[\begin{pmatrix} 0 \\ 0 \end{pmatrix}, \begin{pmatrix} R & C \\ C & Q \end{pmatrix}\right]$; and,

$Z = [1,0,\cdots,0]$ with size $1 \times m$.

From the system above, the updating equations are (i.e. (23)-(25) in the Kalman-filter algorithm (23)-(27)):

$$E_t(RP_t) = E_{t-1}(RP_t) + K_t[fe_t - ZE_{t-1}(RP_t)] \quad \text{(A1)}$$
$$V_t(RP_t) = V_{t-1}(RP_t) - K_t Z V_{t-1}(RP_t) \quad \text{(A2)}$$
$$K_t = V_{t-1}(RP_t) Z^T [Z V_{t-1}(RP_t) Z^T + R]^{-1} \quad \text{(A3)}$$

Consider the conditional mean $E_t(RP_{t+1})$ and divide it into two time-dependence parts: up to $t-1$ and from $t-1$ to $t$.

$$\begin{aligned} E_t(RP_{t+1}) &= E_{t-1}(RP_{t+1}) + E\big(RP_{t+1}|\widetilde{fe}_t\big) \\ &= \Phi E_{t-1}(RP_t) + E\big(RP_{t+1}|\widetilde{fe}_t\big) \end{aligned}$$

where $\widetilde{fe}_t = fe_t - ZE_{t-1}(RP_t) = Z[RP_t - E_{t-1}(RP_t)] + re_{t+1}$.

The joint Gaussian of $RP_{t+1}$ and $\widetilde{fe}_t$ leads to:

$$E\big(RP_{t+1}|\widetilde{fe}_t\big) = E(RP_{t+1}) + \sigma\big(RP_{t+1}, \widetilde{fe}_t\big)\big[\sigma\big(\widetilde{fe}_t\big)\big]^{-1}\big[\widetilde{fe}_t - E\big(\widetilde{fe}_t\big)\big]$$

where $\sigma\big(RP_{t+1}, \widetilde{fe}_t\big) = \sigma\{\Phi(RP_t) + \Theta a_{t+1}, Z[RP_t - E_{t-1}(RP_t)] + re_{t+1}\}$
$= \Phi V_{t-1}(RP_t)Z^T + \Theta C,$
$\sigma\big(\widetilde{fe}_t\big) = ZV_{t-1}(RP_t)Z^T + R.$

As a result,

$$\begin{aligned} E(RP_{t+1}) &= \Phi E_{t-1}(RP_t) + [\Phi V_{t-1}(RP_t)Z^T + \Theta C][ZV_{t-1}(RP_t)Z^T + R]^{-1}\widetilde{fe}_t \\ &\equiv \Phi E_{t-1}(RP_t) + \widetilde{K}\widetilde{fe}_t \qquad \text{(A4)} \end{aligned}$$

Substituting (A1) into (A4), rearranging, and simplifying; then:

$$E(RP_{t+1}) = \Phi E_t(RP_t) + \Theta C[ZV_{t-1}(RP_t)Z^T + R]^{-1}[fe_t - ZE_{t-1}(RP_t)] \quad \text{(A5)}$$

Now, consider the conditional variance $V_t(RP_{t+1})$ with $RP_{t+1} - E(RP_{t+1}) = \Phi[RP_t - E_{t-1}(RP_t)] + \Theta a_{t+1} - E\big(RP_{t+1}|\widetilde{fe}_t\big)$.

$$\begin{aligned} V_t(RP_{t+1}) &= \Phi V_{t-1}(RP_t)\Phi^T + \Theta Q\Theta^T + \widetilde{K}\sigma\big(\widetilde{fe}_t\big)\widetilde{K}^T \\ &\quad -\sigma\big\{\Phi[RP_t - E_{t-1}(RP_t)], \widetilde{K}\widetilde{fe}_t\big\} - \sigma\big\{\widetilde{K}\widetilde{fe}_t, \Phi[RP_t - E_{t-1}(RP_t)]\big\} \\ &\quad -\sigma\big[\Theta a_{t+1}, \widetilde{K}\widetilde{fe}_t\big] - \sigma\big[\widetilde{K}\widetilde{fe}_t, \Theta a_{t+1}\big] \\ &= \Phi V_{t-1}(RP_t)\Phi^T + \Theta Q\Theta^T - \Theta C[ZV_{t-1}(RP_t)Z^T + R]^{-1}\Theta^T C^T \quad \text{(A6)} \end{aligned}$$

Substituting (A2) into (A6), rearranging, and simplifying; then:

$$\begin{aligned} V_t(RP_{t+1}) &= \Phi V_t(RP_t)\Phi^T + \Theta Q\Theta^T - \Theta C[ZV_{t-1}(RP_t)Z^T + R]^{-1}\Theta^T C^T + \\ &\quad - \Phi K_t C^T\Theta^T - \Theta C K_t^T\Phi^T \qquad \text{(A7)} \end{aligned}$$

Eventually, the equations (A1), (A2), (A3), (A5), and (A7) (equivalent to (23)-(27)) collectively establish the Kalman-filter algorithm for the foreign-exchange-risk-premium state-space setting (20)-(22).

## Appendix B: Additional state-space formulations

Apart from $AR(1)$, the additional EViews-state-space specifications probably beneficial for other financial variables are $ARMA(1,1)$ and $MA(1)$ processes.

$ARMA(1,1)$ state space in EViews:

```
@signal fe = sv1 + c(4)*sv2 + [ename = e1]

@state sv1 = c(4)*sv1(-1) + [ename = e2]
@state sv2 = sv1(-1)

@evar var(e1) = exp(c(1))
@evar var(e2) = exp(c(2))
@evar cov(e1, e2) = c(3)
```

$MA(1)$ state space in EViews:

```
@signal fe = sv1 + c(4)*sv2 + [ename = e1]

@state sv1 = [ename = e2]
@state sv2 = sv1(-1)

@evar var(e1) = exp(c(1))
@evar var(e2) = exp(c(2))
@evar cov(e1, e2) = c(3)
```

Let c(3) be 0 if cov(e1,e2) = 0 is assumed.